%% file: cur_paper.tex
\documentclass[10pt,conference]{IEEEtran}

\usepackage{multirow}
\usepackage{makecell}
\usepackage{siunitx}
\usepackage{tabularx}
\usepackage{xcolor}
\usepackage{xspace}
\usepackage{soul}
\usepackage{booktabs}
\usepackage{url}
\usepackage{graphicx}
\usepackage{amsfonts}
\usepackage{hyperref}
\usepackage{cite}
\usepackage{balance}

\newcommand{\ie}{\emph{i.e.,}\xspace}
\newcommand{\eg}{\emph{e.g.,}\xspace}

\newcounter{observation}
\newcommand{\observation}[1]{\refstepcounter{observation}
	\begin{center}
		\framebox{
			\begin{minipage}{0.93\columnwidth}
				{} \textit{#1}
			\end{minipage}
		}
	\end{center}
}

\AtBeginDocument{%
  \providecommand\BibTeX{{%
    \normalfont B\kern-0.5em{\scshape i\kern-0.25em b}\kern-0.8em\TeX}}}

\makeatletter
\newcommand{\linebreakand}{%
  \end{@IEEEauthorhalign}
  \hfill\mbox{}\par
  \mbox{}\hfill\begin{@IEEEauthorhalign}
}
\makeatother

\title{From Commit Message Generation to History-Aware Commit Message Completion}

\author{
\IEEEauthorblockN{Aleksandra Eliseeva}
\IEEEauthorblockA{
    \textit{JetBrains Research}\\
    Republic of Serbia\\
    alexandra.eliseeva@jetbrains.com
}
\and
\IEEEauthorblockN{Yaroslav Sokolov}
\IEEEauthorblockA{
    \textit{JetBrains}\\
    Germany\\
    yaroslav.sokolov@jetbrains.com
}

\and
\IEEEauthorblockN{Egor Bogomolov}
\IEEEauthorblockA{
    \textit{JetBrains Research}\\
    Republic of Cyprus \\
    egor.bogomolov@jetbrains.com
}
\linebreakand
\IEEEauthorblockN{Yaroslav Golubev}
\IEEEauthorblockA{
    \textit{JetBrains Research}\\
    Republic of Serbia \\
    yaroslav.golubev@jetbrains.com
}
\and
\IEEEauthorblockN{Danny Dig}
\IEEEauthorblockA{
    \textit{JetBrains Research}\\ \textit{University of Colorado Boulder}\\
    United States \\
    danny.dig@jetbrains.com
}
\and
\IEEEauthorblockN{Timofey Bryksin}
\IEEEauthorblockA{
    \textit{JetBrains Research}\\
    Republic of Cyprus \\
    timofey.bryksin@jetbrains.com
}
}

\begin{document}
\maketitle

\begin{abstract}

Commit messages are crucial to software development, allowing developers to track changes and collaborate effectively. Despite their utility, most commit messages lack important information since writing high-quality commit messages is tedious and time-consuming. The active research on commit message generation (CMG) has not yet led to wide adoption in practice. We argue that if we could shift the focus from commit message generation to \textit{commit message completion} and use previous \textit{commit history} as additional context, we could significantly improve the quality and the personal nature of the resulting commit messages.

In this paper, we propose and evaluate both of these novel ideas. Since the existing datasets lack historical data, we collect and share a novel dataset called \textit{CommitChronicle}, containing 10.7M commits across 20 programming languages. We use this dataset to evaluate the completion setting and the usefulness of the historical context for state-of-the-art CMG models and GPT-3.5-turbo. Our results show that in some contexts, commit message completion shows better results than generation, and that while in general GPT-3.5-turbo performs worse, it shows potential for long and detailed messages. As for the history, the results show that historical information improves the performance of CMG models in the generation task, and the performance of GPT-3.5-turbo in both generation and completion.

\end{abstract}

\maketitle

\section{Introduction}

Whenever a developer commits their work to a version control system, they can write a short comment in a natural language, called a \textit{commit message}. High-quality commit messages can greatly aid software maintenance, as they provide a human-readable overview of what was changed or why and may ease code review or other activities that require the comprehension of changes~\cite{commit-messages-for-comprehension}. In contrast, poor commit messages can negatively affect software defect proneness~\cite{icse2023quality}. 

Writing a good commit message is tedious and requires extra time and effort from developers. Research shows that a significant amount of commit messages from open source projects lack important information~\cite{what_makes_a_good_msg} or are even empty~\cite{empty_msgs}.

To assist developers, the research community has been working actively on \textit{commit message generation (CMG)}~\cite{deltadoc, changescribe, what_and_why, Jiang2017, loyola2017neural, loyola2018content, ptrmsg,  codisum, coregen, qacom, mcmd, commitbert, bai2021jointly, fira, race}, which is defined as: given the changes made in a commit, generate an appropriate message. 
Despite numerous technical advances in the field, we still do not have wide adoption of CMG in practice, as several practical aspects are not yet solved. Researchers~\cite{Jiang2017, nngen, qacom} found out that over 50\% of messages from existing CMG approaches were inadequate, \textit{i.e.}, semantically irrelevant to the reference messages.

\begin{figure}
    \centering
    \includegraphics[width=0.49\textwidth]{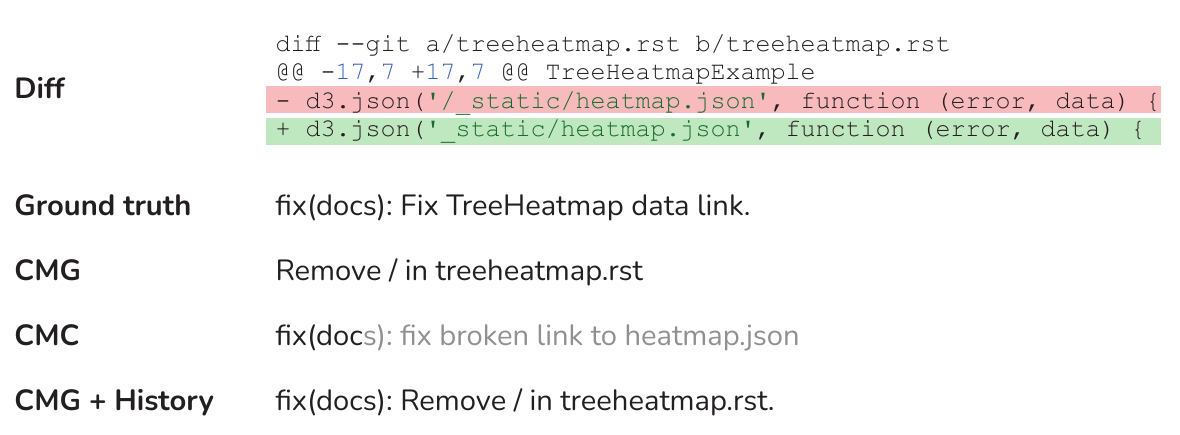}
    \vspace{-0.7cm}
    \caption{Motivating example for our two ideas for personalizing commit message generation. CMG = standard commit message generation; CMC = commit message completion; CMG + history = commit message generation with commit message history as additional context.}
    \label{fig:motivation}
    \vspace{-0.6cm}
\end{figure}

In this paper, we propose two novel approaches to improve the quality of generated commit messages from the standpoint of their personalization: \emph{commit message completion} instead of generation, and \emph{taking the user's previous commit messages into account}. To the best of our knowledge, neither of these approaches has been tested in this context before.

First, we rephrase the problem of commit message generation into \textit{commit message completion}. This way, the already-typed prefix of the commit message helps the model suggest the following relevant tokens.
Completion systems for both natural languages~\cite{gmailcompose} and programming languages~\cite{intellicode} are now widespread and overall well accepted by end users. In addition, practitioners explicitly name completion as a way they might prefer to use CMG approaches~\cite{corec}. We hypothesize that a prefix might guide the existing approaches towards more applicable predictions and improved adherence to project conventions, as illustrated in Figure~\ref{fig:motivation}.

Second, to aid personalization, we take into account syntactic and stylistic conventions of a particular project or user by \textit{considering the history of their commits} as a part of the model's input. Figure~\ref{fig:motivation} also illustrates this. Such an approach requires two things: \textbf{(a)} we must have this \textit{history} saved in the training dataset, and \textbf{(b)} such a dataset must be not only large but also \textit{diverse} for the models to learn various possible conventions and realistic commit messages. 

We start our work by studying existing CMG datasets. Unfortunately, all of them have the same crucial shortcomings: no saved history and restrictive data filtering.
To mitigate this issue, we build a large-scale multilingual dataset that incorporates the best practices from previous works while avoiding the use of filters restricting the representativeness of the data. The dataset is called \textit{CommitChronicle} and contains ${\sim}10.7$M commits in $20$ programming languages from ${\sim}12$K repositories with permissive licenses. 
To the best of our knowledge, our dataset is the only one that both provides author metadata and keeps commit history close to the origins. 

\begin{figure*}
    \centering
    \includegraphics[width=0.99\textwidth]{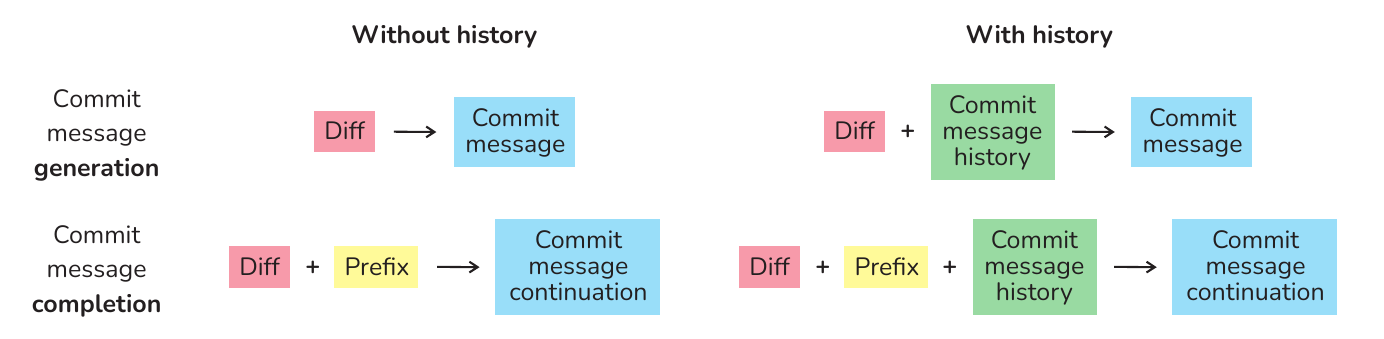}
    \vspace{-0.3cm}
    \caption{Overview of the four configurations considered in our study and their contents.}
    \label{fig:overview}
    \vspace{-0.3cm}
\end{figure*}

Since we suggest two separate ideas---reformulating commit message generation to completion and using commit message history---we experiment with all four possible configurations, as shown in Figure~\ref{fig:overview}.

To evaluate the completion setting, we study how different models perform in it. 
From CMG models, we experiment with CodeT5~\cite{codet5}, RACE~\cite{race}, and CodeReviewer~\cite{codereviewer}. Our research demonstrates that metrics scores generally increase when the bigger part of the original message is passed to the context. This finding suggests that completion could be an effective way to apply existing approaches. 
We also experiment with a large language model GPT-3.5-turbo (also known as ChatGPT) with a simple and straightforward prompt, which demonstrates generally worse results than dedicated CMG models, however, shows potential for longer and more detailed messages. Our findings suggest that the completion setting is significantly easier in practice: for the best model, the average \textit{B-Norm} value grows from 16.9 in the generation setting to 27.2 when the user already typed half of the message.

We evaluate our second idea of considering the commit history and show that the history improves the performance of CMG models for generation (\textit{B-Norm} improves from 15.3 to 16.9), however, the results are conflicting for completion. As for GPT-3.5-turbo, the results improve in both settings. Finally, we study the impact of restrictive data filters by comparing the models on commits that pass all the restrictions and those that do not.
Our comparison shows that the results are much better for commits passing all the restrictions than for the commits that pass neither of the filters (\textit{B-Norm} five times higher), indicating that the existing datasets might inflate the results and fail to account for a lot of in-the-wild commits.

Overall, our results show that both ideas---commit message completion and taking into account the history of commits---show potential in specific scenarios and require further research. 
This paper makes the following contributions:

\begin{itemize}

    \item A \textbf{novel reframing} of commit message generation into completion, its formulation, and an evaluation setup.
      
    \item An \textbf{approach} of appending the history of commit messages into the model's input for the tasks of commit message generation and completion.
   
    \item A comprehensive \textbf{analysis} of existing CMG datasets from the standpoint of them keeping commit message history and the restrictiveness of filters, which shows that they have shortcomings from both points of view.
    
    \item A \textbf{diverse dataset} called \textit{CommitChronicle} with ${\sim}10.7$ million commits in $20$ programming languages, which preserves information necessary for utilizing commit history. We make the dataset publicly available~\cite{dataset}.    
    
    \item A \textbf{comparison} of three state-of-the-art CMG models (CodeT5, RACE, and CodeReviewer) and GPT-3.5-turbo from multiple points of view, including generation against completion and history against no history. Our main results indicate that in some cases, completion is simpler than generation, and using history improves the performance of models. All the code for our models and experiments is available online~\cite{artifacts}.
   
\end{itemize}

\section{Background}
\label{sec:background}

In this section, we give a gentle introduction to the main relevant concepts: commit message generation, completion as used in various domains, and the current uses of large language models (LLMs) in software engineering.

\textbf{Commit message generation}.
Modern CMG approaches can be broadly categorized into three groups: language modeling-based, retrieval-based, and hybrid.

CMG may be considered as a task of sequence-to-sequence \textit{language modeling}, analogous to neural machine translation~\cite{Jiang2017} or summarization~\cite{loyola2017neural}.  
In this setting, both input and output are represented as sequences of tokens. During the training phase, given a source sequence $\mathbf{x} = (x_1, \dots, x_n)$ 
and a target sequence 
$\mathbf{y} = (y_1, \dots, y_m)$,
the language model learns a probability distribution
$p(y_1, \dots, y_m | x_1, \dots, x_n)$
by optimizing the loss function.

Most language modeling-based CMG approaches follow the encoder-decoder structure~\cite{Jiang2017, loyola2017neural, loyola2018content, ptrmsg, coregen, commitbert}. The encoder receives a source sequence $\mathbf{x}$
and produces a vector representation $\mathbf{h} \in \mathbb{R}^d$, where $d$ is a hyperparameter. 
The decoder receives $\mathbf{h}$ and produces an output sequence 
$\mathbf{y}$
in an autoregressive manner, \textit{i.e.}, utilizing only previous tokens for each position.  
Specific neural network architectures considered in prior works include Recurrent Neural Network (RNN) and its modifications~\cite{Jiang2017, loyola2017neural, loyola2018content, ptrmsg, codisum},  Transformer~\cite{coregen, commitbert, race}, and Graph Neural Network (GNN)~\cite{fira}.
 
Unlike language modelling, \textit{retrieval-based} approaches find a commit with the most similar diff in the training set and return a corresponding message. The first such approach is \textit{NNGen}~\cite{nngen}: it uses bag-of-words to represent diffs, retrieves $k$ most similar commits based on cosine similarity, and returns the commit message based on the BLEU score~\cite{bleu} between diffs. Further enhancements to the framework include limiting retrieval to commits from the same repository~\cite{etemadi2020relevance}, replacing the metric~\cite{mcmd}, and considering more complex approaches for representing diffs~\cite{cc2vec, changedoc}. Retrieval approaches rely on the presence of similar messages in the training set and may fail due to the unique code identifiers in commit messages.

Finally, \textit{hybrid} approaches aim to combine the advantages of both language modeling and retrieval. One option is to choose between retrieved and generated messages~\cite{atom} and the other is to employ the retrieved commits inside the language model to improve the generation quality~\cite{corec, race}. At the moment of writing, the recently proposed hybrid approach \textit{RACE}~\cite{race} is state of the art in CMG.

Despite all these different approaches, the question of their practical usefulness remains open. In several studies, over 50\% of generated messages were rated as low-quality by human experts~\cite{Jiang2017, nngen, qacom}.

\textbf{Completion}. 
The completion task might be viewed as a simpler version of generation. Rather than generating the full message, completion systems aim to assist a user by suggesting relevant continuation when they start typing. This framework is widely used in practice for diverse domains: writing emails~\cite{gmailcompose} or code completion in IDEs~\cite{intellicode, bibaev2022all}. Popular approaches employ language modelling~\cite{gmailcompose, intellicode, mastropaolo2021empirical} or ranking, however, the latter is suitable for domains where a limited amount of candidates can be obtained, \textit{e.g.}, code completion~\cite{svyatkovskiy2021fast}.

\textbf{LLMs in software engineering}. 
LLMs are language models scaled up to billions of parameters and billions of tokens in a training set. The majority of recent LLMs are based on the Transformer decoder~\cite{transformer} architecture. A particularly interesting property of LLMs is the in-context learning, \ie the ability to solve downstream tasks in a few-shot setting (with just several examples passed as the context) or even zero-shot setting (without any examples, only an instruction passed as the context), without actually updating model weights~\cite{gpt3}. 

As for the capabilities of LLMs in the software engineering domain, a notable example is the code-specific model \textit{Codex} from OpenAI~\cite{codex}. It was shown to achieve good performance on a variety of software engineering tasks, including both code generation~\cite{codex, codex-testgen, codex-bugrepro, codex-repair1} and natural language generation~\cite{codex-doc, codex-codesum, codex-rootcause}. 
Recently, two general-purpose LLMs were introduced by OpenAI --- \textit{GPT-3.5-turbo} (ChatGPT) and \textit{GPT-4}. An overview of their capabilities can be found in the GPT-4 technical report~\cite{gpt4} or in the survey from Liu et al.~\cite{chatgpt_gpt4_survey}.

\section{Commit Message Completion}
\label{sec:cmc}

Our first approach for improving the practical usefulness of existing solutions is moving from commit message generation to commit message completion. In this section, we formally define the task and describe the specifics of how it works with different model architectures.

\textbf{Task definition}.
Commit message completion can be formulated as follows: given the prefix of a commit message and the \textit{commit context}, generate a suitable continuation for the commit message. In practice, the length of this continuation may vary based on the quality and performance of the underlying approach and other factors.
We follow the existing CMG works~\cite{Jiang2017, ptrmsg, nngen, mcmd} and use diff (\ie changes made in the commit) as the main commit context. 
Figure~\ref{fig:overview} presents the high-level overview for the commit message completion task with \textit{commit contexts} that we consider.
We focus on the language modeling approach to completion because it is widely adopted for natural language sequences~\cite{gmailcompose, mastropaolo2021empirical}.

\textbf{Input representation}.
In this study, we do not seek to utilize code structure and leave this to future research. We have two key observations to motivate our decision. Firstly, developers use version control systems not only for source code files but also for various related files (\textit{e.g.}, configurations and READMEs), which do not necessarily follow a formal structure.
Secondly, most of the existing approaches that utilize code structure only support Java~\cite{atom, codisum, fira}, and it would require substantial effort to support more languages. To maintain the diversity necessary for the purposes of our study, we consider it essential to employ a dataset covering multiple programming languages and therefore use a plain textual representation of changes made in each commit --- an output of the \texttt{git diff} command. 

\textbf{Encoder-decoder approach}.
Language modeling-based CMG approaches can be directly applied to the completion task. During training, we use the commit diff as the source sequence $\mathbf{x}$, and the ground truth commit message as the target sequence $\mathbf{y}$. 
During inference, the model takes a commit diff $\mathbf{x}$ and a prefix for a commit message $\mathbf{p_m}$. We pass $\mathbf{p_m}$ to the decoder since the output is expected to be its continuation.
Optionally, we may include an additional \textit{commit context} $\mathbf{c_m}$ (\textit{e.g.}, containing this author's commit history). During training, we concatenate it with the ground truth commit message and build a target sequence $\mathbf{y_{c_m}} = (\mathbf{c_m}, [SEP], \mathbf{y})$, where $[SEP]$ is a special token. We compute and propagate loss only for the message $\mathbf{y}$.
During inference, we concatenate it with the prefix $\mathbf{p_m}$ in the same manner.

\textbf{LLM approach}.
In contrast with the CMG approaches, LLMs mostly follow the decoder-only architecture instead of the encoder-decoder. Also, we focus solely on evaluating the in-context learning abilities of LLMs, \ie we do not consider fine-tuning. Hence, several differences arise. The input for an LLM is a single sequence of tokens, often referred to as \textit{prompt}. LLM outputs a continuation for the given prompt. There are many techniques for prompt engineering (\textit{i.e.}, the process of choosing the best-performing prompt)~\cite{white2023chatgpt}, however, we focus on a simple zero-shot setting as a baseline. Specifically, we provide the model only with a simple instruction to complete the given commit message prefix based on commit diff and, optionally, additional \textit{commit context}.

\input{tables/04-datasets-statistics}

\section{Commit History \& Diversity of Data}
\label{section:Limitations of Existing Datasets}

The second important improvement is utilizing the previous commit messages, as shown in Figure~\ref{fig:overview}, which requires training the models on the appropriate data. In a recent overview, Tao et al.~\cite{mcmd} highlighted that the majority of the available CMG datasets suffer from at least one of the following limitations: \textbf{(a)} only the Java language; \textbf{(b)} small scale of 20,000--100,000 commits; \textbf{(c)} limited information about each commit (hence, no way to trace back to the original commit on GitHub).

To mitigate these issues, Tao et al.~\cite{mcmd} built a novel dataset called MCMD. However, we argue that there are two more important limitations in the existing CMG datasets: \textit{significant tampering with the original commit history} and \textit{restrictive data filtering}. The former undermines the validity of experiments with commit history and limits the possibility of taking into account individual characteristics of specific developers and projects, while the latter impedes the ability to learn potential conventions among a diverse range of commits and to evaluate on them. Let us now describe these limitations.

\textbf{Tampering with the original commit history}. 
To the best of our knowledge, at the time of publishing, our work is the first to explore the commit message history as an additional source of information for generating commit messages. We observe that preserving the original commit history was out of scope for the majority of existing published CMG datasets. This shortcoming manifests itself in one of the following ways: \textbf{(a)} preserving only a small fraction of the original set of commits due to strict filters~\cite{Jiang2017, nngen}; \textbf{(b)} performing train/validation/test split randomly, not by authors or by projects~\cite{Jiang2017, nngen, codisum, atom, ptrmsg, corec, commitbert}; \textbf{(c)} downsampling (\ie selecting a subset) from the original set of commits randomly, without considering authors or projects~\cite{mcmd, commitbert}.

\textbf{Restrictive data filtering}.
Even when it comes to individual commit messages, existing works usually apply restrictive filterings that relate both to the messages themselves and the diffs. Notice that we do not count as restrictive filtering those steps that aim to lower the number of automatically generated examples (\textit{e.g.}, dropping merge or revert commits). Let us highlight the most common and important filters that researchers used in previous works.
First, there are filters that relate to \textit{commit messages}:

--- \textit{First Sentence}. Developers often use the first sentence of a commit message as a concise summary of the entire message~\cite{Jiang2017}. Many CMG papers opt to extract the first sentence from commit messages as a target sequence.
    
--- \textit{Message Structure}. Commit messages from open-source projects vary drastically in terms of writing styles, and some filters are employed by CMG papers to restrict this variety. A notable example is the Verb-Direct Object (V-DO) filter, which only allows messages that start with a Verb followed by a Direct Object clause~\cite{Jiang2017} (\textit{e.g.}, \textit{refactor code}, but not \textit{minor refactoring}). Another option is filtering out commit messages that do not begin with one of the curated verbs~\cite{commitbert}.

--- \textit{Message Length}. This filter is usually targeting the number of tokens in commit messages.

Other filters target \textit{commit diffs}:

--- \textit{Only Code}. Some studies only consider commits that only modify source code files (\textit{i.e.}, \texttt{.java} for Java).
    
--- \textit{Diff Length}. This filter is usually targeting the number of tokens in diffs. Other variations include the number of changed files or chunks (changed lines grouped together).

We studied the corresponding papers and replication packages of existing CMG datasets and provide the resulting statistics on the usage of each filter in Table~\ref{tab:filters3}. \textit{First Sentence}, \textit{Message Structure}, \textit{Message Length}, and \textit{Diff Length} are used in more than half of existing CMG datasets. The only dataset that does not employ any of these filters is the one from the work of Loyola et al.~\cite{loyola2017neural}, however, it only contains commits from 12 specific projects. Also, we note that the dataset from the work of Jiang et al.~\cite{Jiang2017} and its filtered version NNGen~\cite{nngen} are the most common for evaluation of CMG models~\cite{ptrmsg, cc2vec, etemadi2020relevance, coregen}, and several subsequent works employ the same processing pipeline for their datasets~\cite{ptrmsg, corec}.

There is evidence for the restrictiveness of these filters. Jiang et al.~\cite{Jiang2017dataset} explored 1.6 million commit messages from top 1,000 Java projects, and their findings show that 53\% of messages do not follow the Verb-Direct Object structure and 18\% of the messages have more than one sentence. In the later work, Jiang et al.~\cite{Jiang2017} employ the \textit{Message Length} and \textit{Diff Length} filters for $30$ and $100$ tokens, respectively. This filtered 1.8M commits into 75K commits (a reduction of~96\%).

Drastic reductions used in previous research datasets impede the ability to study the history-based personalization of commit messages. In our work, we aim to bridge this gap by collecting a new large-scale, multilingual, history-aware, diverse dataset called \textit{CommitChronicle}, and studying whether the history and filterings significantly influence the results.

\section{The CommitChronicle Dataset}
\label{sec:Dataset}

\subsection{Data Collection}

\textbf{Choosing repositories}.
As our source of information, we chose GitHub, a large platform for hosting software projects. We used the GitHub Search tool~\cite{github_search} on January 25th, 2023 to select specific repositories for subsequent data mining. 
To filter only mature projects, we set the inclusion criteria based on the existing guidelines~\cite{kalliamvakou2014promises},   similar to other works in SE research~\cite{wang2021pynose, grotov2022large}: 50+ stars, 10+ contributors, 1000+ commits, created at least two years ago, has a permissive license (Apache-2.0, MIT, BSD-3-Clause), not a fork. In total, we obtained $12.4$k projects fitting our criteria.

\textbf{Collection process}. The data collection took place on February 9th, 2023. We used PyDriller~\cite{Spadini2018} to collect commits. We collected all non-merge commits made after January 1st, 2017, opting to avoid earlier commits since they might be less relevant. In order to fit into reasonable resources for data processing, we also set the upper limit on the number of changed lines in a single commit to 10,000. In total, we obtained $27.4$M commits. 

\subsection{Data Processing}

\textbf{Splitting by projects}.
To evaluate the models on the previously unseen projects and to avoid breaking the commit history, we split the data into the train, validation, and test sets by repositories with the 80\%/10\%/10\% ratio adopted in previous work~\cite{mcmd}.

\textbf{Filtering outliers}.
The main purpose of this stage is to drop examples that are both highly atypical and require a lot of time and memory to process. We calculated percentiles for the number of tokens (obtained via simple tokenization by whitespaces), number of characters, and number of modified files. We dropped examples out of the [5\%, 95\%] percentile range. We provide the exact percentile values in our online appendix~\cite{artifacts}.
This resulted in $21$M commits, with $6.4$M commits dropped ($23.32\%$ of commits from the initial step).
    
\textbf{Commit message processing}.
Unlike most of the existing CMG datasets, we refrained from processing that would restrict the diversity of commit messages. Nevertheless, we adopted the best practices to filter out automatically generated or irrelevant commit messages, including messages with non-ASCII symbols~\cite{Jiang2017dataset}, trivial messages~\cite{nngen}, and merge and revert messages~\cite{Jiang2017}.
In addition, we identified and removed project-specific content from the messages, including URLs, emails, and references to issues or pull requests~\cite{filter_messages}.
We provide all the regular expressions in our online appendix~\cite{artifacts}. 
In total, this resulted in $19.2$M commits, with $1.8$M commits dropped ($8.81\%$ of commits from the previous step).

\textbf{Commit diff processing}.
We store diffs as a list of file modifications, where each modification includes type (modifying file, creating file, deleting file, etc.), path to file before and after commit, and diff as obtained from Git. For diffs, we merge several consecutive whitespaces to save up disk space. Also, we drop commits with empty diffs. In total, this resulted in $18.6$M commits, with $591$K commits dropped ($3.08\%$ of commits from the previous step).
    
\textbf{Deduplication}.
To ensure that our evaluation results are not inflated, we conducted exact hash deduplication of our dataset. Specifically, we group commits that share the same MD5 hash for their messages or diffs, and keep a single commit instance from each group. In total, we found and dropped $4.6$M duplicate commits ($24.59\%$ of commits from the previous step). This resulted in $14.0$M commits.

\textbf{Dropping commits from overlapping authors}.
To prevent any overlap between commit authors in the training and evaluation sets, we removed all commits from authors who were present in either the validation or test set from the training set. Out of $488.5$K authors in the training set, we identified $54.0$K overlapping authors ($11.07\%$). In total, this resulted in $10.8$M commits, with $3.2$M commits dropped ($22.55$\% of commits from the previous step).

\textbf{Dropping commits from bots}.
It is important to note that open-source projects frequently employ software bots to automate certain activities~\cite{bots_bodegha, bots_accuracy, bots_biman, bots_bothunter}.
To further lower the number of automatically generated commits, we drop all commits from the authors that either are present in existing bot datasets~\cite{bots_bodegha, bots_bodegic, bots_bothunter} or have the suffix ``bot" in their names~\cite{bots_accuracy}.
Specifically, we identified $902$ out of $603$K authors as bots and dropped $165.8$K commits ($1.52\%$ of commits from the previous step). In total, this resulted in $10.7$M commits, which is the final number of commits in our dataset.

\textbf{Name disambiguation}.
After all the processing, we replace the authors' names and emails with unique identifiers to prevent personal information disclosure. Since the same user might appear under several combinations of name and email (\textit{e.g.}, work email and personal email), we employ a name disambiguation tool \textit{gambit}~\cite{gambit} to obtain identifiers and merge authors together. \textit{gambit} was shown to achieve near-perfect results for the authors from the Gnome GTK project.  
\textit{gambit} calculates text similarity metrics for all pairs of authors: since processing all author pairs in our large dataset would be infeasible, we limited name disambiguation to the authors committing to the same repository. Consequently, if some author committed to several repositories, they will have different identifiers for each repository. Due to this limitation, we performed two preceding steps related to commit authors before name disambiguation rather than after it.

\subsection{Dataset Overview}

\textbf{General statistics}.
Table~\ref{tab:final_statistics} presents the resulting number of commits in our dataset. Notice that we performed the splitting by repositories, which resulted in the uneven distribution in terms of commits, especially for low-resource languages (\textit{e.g.}, Nix or Groovy). After all processing, we retained $10.7$M of the initial $27.4$M commits ($38.98$\%), with the most restrictive stages being \textit{Filtering outliers} and \textit{Deduplication}, which are necessary for data cleaning. In terms of projects, we retained $11.9$k of the initial $12.4$k projects ($95.97$\%). Following the MCMD dataset~\cite{mcmd}, we include not only diffs and messages but also rich metadata about each commit, including authors and timestamps necessary for experiments with the commit history, as well as repository names and commit hashes, which allow researchers to gather additional information if needed. 

\input{tables/05-our-dataset}

\textbf{Filters and commit history statistics}.
In Section~\ref{section:Limitations of Existing Datasets}, we observed that many existing CMG datasets employ restrictive filters. \textit{CommitChronicle} is large-scale and multilingual, hence, it is representative to assess the degree of restrictiveness for each filter that other researchers used previously. 
We implement the most frequent filters and calculate the number of affected commits for each of them. We provide the implementation for each filter in our online appendix~\cite{artifacts}.
Our findings show that \textit{Diff Length $\leq 100$ Tokens} and \textit{Verb-Direct Object} are the most restrtictive filters: in our dataset, we would  have excluded $84\%$ and $64\%$ commits, respectively, if we employed these filters. With \textit{First Sentence}, $17\%$ commits would be filtered out. Finally, \textit{Message Length $\leq 30$ Tokens} would only affect $7\%$ commits, which suggests that commit messages tend to be concise in open source projects.

Another shortcoming of existing CMG datasets outlined in Section~\ref{section:Limitations of Existing Datasets} that makes them unsuitable for experiments with commit history is that they either do not provide the required metadata or tamper with the original commit history. In contrast, we provide the authors' identifiers and timestamps for all commits in our dataset.
As for the commit histories, even after cleaning the data to ensure its quality, on average, we still retain $67$\% of commits in each history, which we consider enough for meaningful experiments. 

\section{Methodology}
\label{sec:methodology}

We evaluate our two ideas for helping developers write high-quality commit messages: commit message completion instead of generation, as well as training and evaluating the models on the newly-collected history-aware and diverse \textit{CommitChronicle} dataset. We conduct experiments with all four possible configurations shown in Figure~\ref{fig:overview}.
To thoroughly cover our two major ideas, we designed four research questions:

\begin{itemize}
    \item \textbf{RQ A1}. How do state-of-the-art CMG approaches perform in the completion setting?

    \item \textbf{RQ A2}. How do LLMs perform in comparison with state-of-the-art CMG approaches?

    \item \textbf{RQ B1}. How does using commit message history as an additional input affect the models' quality?

    \item \textbf{RQ B2}. How do state-of-the-art CMG approaches perform with and without common data filtering steps?
\end{itemize}

In this section, we describe in detail the methodology for answering these research questions.

\textbf{Models}.
Since our dataset includes many programming languages rather than focusing on a single one, this makes it infeasible to adapt the approaches that utilize code structure. Also, retrieval approaches are not directly applicable to the completion task. Therefore, we focus on language modeling or hybrid approaches that do not require structural information.

Specifically, we consider approaches that were shown to achieve the best performance in a recent CMG study~\cite{race}. We also expand the list with a recent model that was shown to achieve superior performance on a variety of tasks related to commit diffs, as well as a powerful LLM.

\textit{CodeT5}~\cite{codet5} is a variation of the sequence-to-sequence language model T5~\cite{t5} that was pretrained on a large amount of source code with source code-specific pretraining objectives. It was shown to achieve good performance on a variety of downstream tasks, including CMG~\cite{race}.

\textit{RACE}~\cite{race} is a hybrid CMG approach that utilizes similar commit messages and commit diffs to improve the quality of a language model. RACE with CodeT5 as a backbone is state of the art in CMG at the time of writing. 
We observed that the retrieval implementation in the RACE replication package is too demanding for the scale of our dataset both in terms of memory and time complexity. Hence, we reimplemented RACE with the retrieval powered by the open-source Approximate Nearest Neighbors (ANN) tool \textit{annoy}~\cite{annoy}.

\textit{CodeReviewer}~\cite{codereviewer} is a variation of CodeT5 that was further pretrained on a large-scale dataset of diffs and code review comments from GitHub. While code review comments and commit messages serve different purposes, we still consider CodeReviewer relevant to the CMG task. In particular, its pretraining includes objectives for understanding commit diffs, which has the potential to bring value for the CMG task.

\textit{GPT-3.5-turbo} is a recent general-purpose LLM from \mbox{OpenAI}. It is officially recommended to be used instead of Codex, as Codex has been deprecated as of March 2023~\cite{deprecation}. We only aim to obtain a reasonable baseline of possible CMG performance of LLMs, hence, we leave the experiments with the more sophisticated GPT-4 model to future research.

\textbf{Training and evaluation setting}.
For all CMG models (\ie all models except for GPT-3.5-turbo), we choose the \textit{base} configuration, since it is the only configuration available for CodeReviewer, and it was shown to be superior over \textit{small} configuration~\cite{race}. 
For all models, we set the maximum source length and target length to \num{512} tokens. 

We use the mixed precision for CMG models to accelerate both training and evaluation. We follow the hyperparameters setting from the most recent CMG study~\cite{race}. Specifically, we use the AdamW optimizer with \num{2e-5} peak learning rate, no weight decay, and a linear warmup strategy with \num{100} warm-up steps. The training is conducted either on a single NVIDIA TITAN RTX GPU or on 4 NVIDIA T4 GPUs via the Distributed Data Parallel (DDP) framework. Effective batch size is set to \num{32} for all runs, with gradient accumulation due to the limited GPU memory when needed. Due to the large-scale nature of our dataset, we train all the models for \num{1} epoch only, which makes $7.6$M examples and around $3$B tokens. During the evaluation, we use beam search with width \num{5} as the decoding strategy, set the maximum number of generated tokens to \num{15}, and allow early stopping when a special end-of-sequence token is produced. 

For GPT-3.5-turbo, we access the model through the official OpenAI API.
OpenAI API provides several configurable parameters for completion endpoints. We set the temperature to 
\num{0.8} and top-p to \num{0.95}, as it was done in previous works~\cite{codex, codex-repair1}. To match our setting for the CMG approaches, we set the maximum generation length to \num{15} tokens and truncate the diffs to \num{512} tokens when constructing prompts.

\textbf{Models' input}.
We mimic a real-world completion scenario by passing the first $X\%$ of characters of each commit message into the context. We experiment with several values of $X$ --- $0\%$ (generation setting), $25\%$ and $50\%$ (completion settings).

For GPT-3.5-turbo, we instruct it with zero-shot prompting. Note that it is a chat model, so the expected data format differs from completion models. We rely on official guidelines to properly define the input in the required format. We provide all the prompts in our online appendix~\cite{artifacts}.

All the CMG models that we consider utilize Byte-Pair Encoding (BPE)~\cite{bpe} subword tokenization algorithm. Hence, for completion, we face an issue with the tokenization of the last incomplete word in the prefix outlined by Popov et al.~\cite{popov-etal-2021-time}. To mitigate it, we remove the last incomplete word from the model input and restrict the beam search to produce outputs consistent with the removed word part.
For GPT-3.5-turbo in the completion setting, we explicitly instruct it to continue given prefixes rather than generate messages from scratch.

We also experiment with commit message history as additional context, following the approach we described in Section~\ref{sec:cmc}. 
For CMG models, we use as many previous commit messages as fit into the context when concatenated with the ground truth message, separating historical messages with a special token $[SEP]$. For GPT-3.5-turbo, we opt for a simpler setting due to its availability through paid API and extend the prompts only with a single previous message from the commit histories of the respective authors within the same repository. Hence, we probe a baseline for what GPT-3.5-turbo may achieve with commit message history.
 
\textbf{Evaluation metrics}.
We employ three metrics commonly used to evaluate the quality of generated commit messages:

\begin{itemize}

\item \textit{B-Norm} is a version of BLEU~\cite{bleu} that was shown to be the most in line with human judgment on the quality of commit messages by Tao et al.~\cite{mcmd}. BLEU is a metric based on the precision in terms of overlapping n-grams (\ie contiguous sequences of \textit{n} words) between generated and reference sequences. It is used in most of prior commit message generation works~\cite{Jiang2017, nngen, mcmd}.

\item \textit{Edit Similarity} (\textit{EdSim}) is a metric based on the Levenshtein distance~\cite{levenshtein} between the predicted and reference texts. It is suitable for evaluating completion systems because users can accept slightly wrong suggestions as long as not too many edits are required~\cite{intellicode}.

\item \textit{ExactMatch} (\textit{EM}) measures the percentage of predicted sequences that exactly match the reference text.

\end{itemize}

We not only report the metrics between full predictions and targets but also investigate how the quality evolves with the number of generated tokens. In addition to the metrics for full sequences, we calculate metric values between prefixes of predictions and targets. For \textit{ExactMatch}, we report the values for prefixes of $1$ and $2$ tokens, since perfect predictions quickly deteriorate to near-zero values. For \textit{B-Norm}, we report metrics for prefixes of $4$--$10$  tokens, since the implementation we use expects $4$-grams to contribute to the final score. For \textit{EdSim}, we report metrics for $1$--$10$ tokens.

\textbf{Data}.
We use the \textit{CommitChronicle} dataset described in detail in Section~\ref{sec:Dataset}. However, we observe that generating predictions for the whole test set with $1.5$ million commits would require an implausible amount of time. We obtain a subsample $CMG_{test}$  as follows: \textbf{(a)} we exclude the repositories with the number of commits more than 95\% percentile (4,161 commits); \textbf{(b)} we downsample repositories for frequent languages to $17$ repositories, to cap the number of commits at around $20$K per language. We do not downsample less represented languages from Table~\ref{tab:final_statistics}, and keep a reasonable trade-off between the data quantity and diversity. In total, $CMG_{test}$ contains $204,336$ commits.

Moreover, since we focus on LLMs available through paid API, we select a smaller yet subsample for all related experiments to keep the costs reasonable. Specifically, we obtain $LLM_{test}$ by randomly selecting $10$ authors with \mbox{$10$--$50$} commits for each programming language from $CMG_{test}$. In total, $LLM_{test}$ contains $4,025$ commits. Note that many previous works used comparable or even smaller datasets when experimenting with LLMs~\cite{codex, codex-codesum, codex-rootcause, codex-repair1}.

\input{tables/07-1-generation-completion}

\section{Results \& Discussion}
\label{sec:results}

Let us now describe the results of our experiments. We use paired bootstrap resampling~\cite{paired-bootstrap-resampling} with $99\%$ confidence level to test statistical significance across different models and settings, \ie to compare a pair of models, we randomly sample with replacement, generating a sample of the same size as the test set, and compute metrics for both models on this new sample; we repeat the process $1,000$ times and declare a winner only if it outperforms the other model in $99$\% of cases.
Due to the lack of space, we only share several plots with metrics between prefixes related to particularly interesting findings. 
We provide all the remaining plots in our online appendix~\cite{artifacts}.

\subsection{Commit Message Completion}

\textbf{RQ A1: How do state-of-the-art CMG approaches perform in a completion setting?}
We present the metrics for the CMG approaches in Table~\ref{tab:main-full-metrics}. First, we observe that the values of metrics tend to grow as we increase the context ratio (\textit{i.e.}, go from generation to completion), especially in \textit{ExactMatch} and \textit{B-Norm} metrics. 
For example, for CodeT5, \textit{ExactMatch@1} grows from $10.90$ (line 4, 0\%) to $45.35$ (line 10, 25\%), to $49.94$ (line 16, 50\%). We confirm that improvements for all models and metrics are statistically significant. 
Hence, to some extent, we confirm the previous finding~\cite{mastropaolo2021empirical} --- completion is a simpler task than generation and might be a good choice to employ existing models to bring value to actual users. 

\begin{figure}
    \centering
    \includegraphics[width=0.49\textwidth]{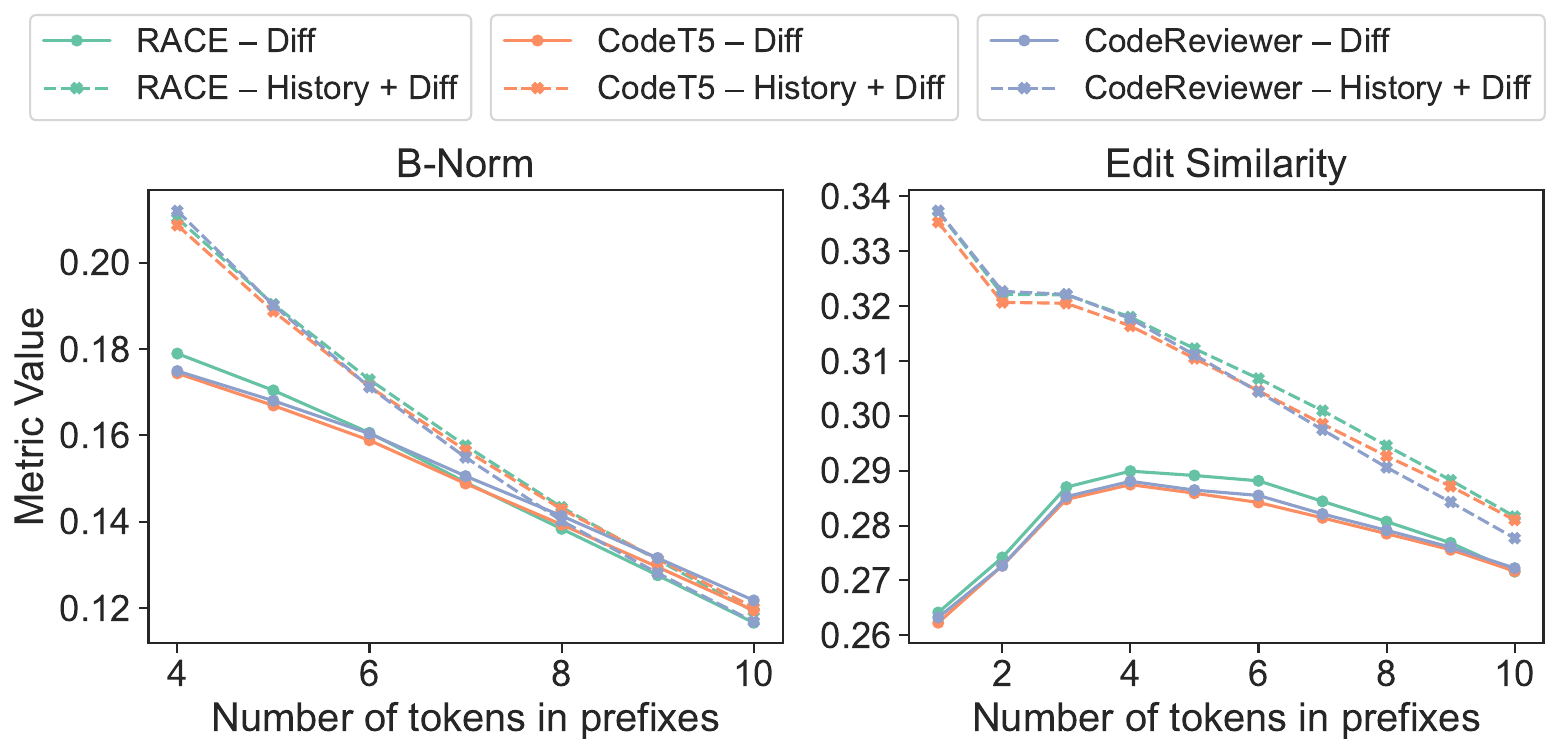}
    \vspace{-0.5cm}
    \caption{Metrics between prefixes on $CMG_{test}$ for 0\% context ratio (generation setting) for CMG approaches. Dashed lines correspond to models with history.}
    \label{fig:prefix_metrics_0}
\end{figure}

\begin{figure}
    \centering
    \includegraphics[width=0.49\textwidth]{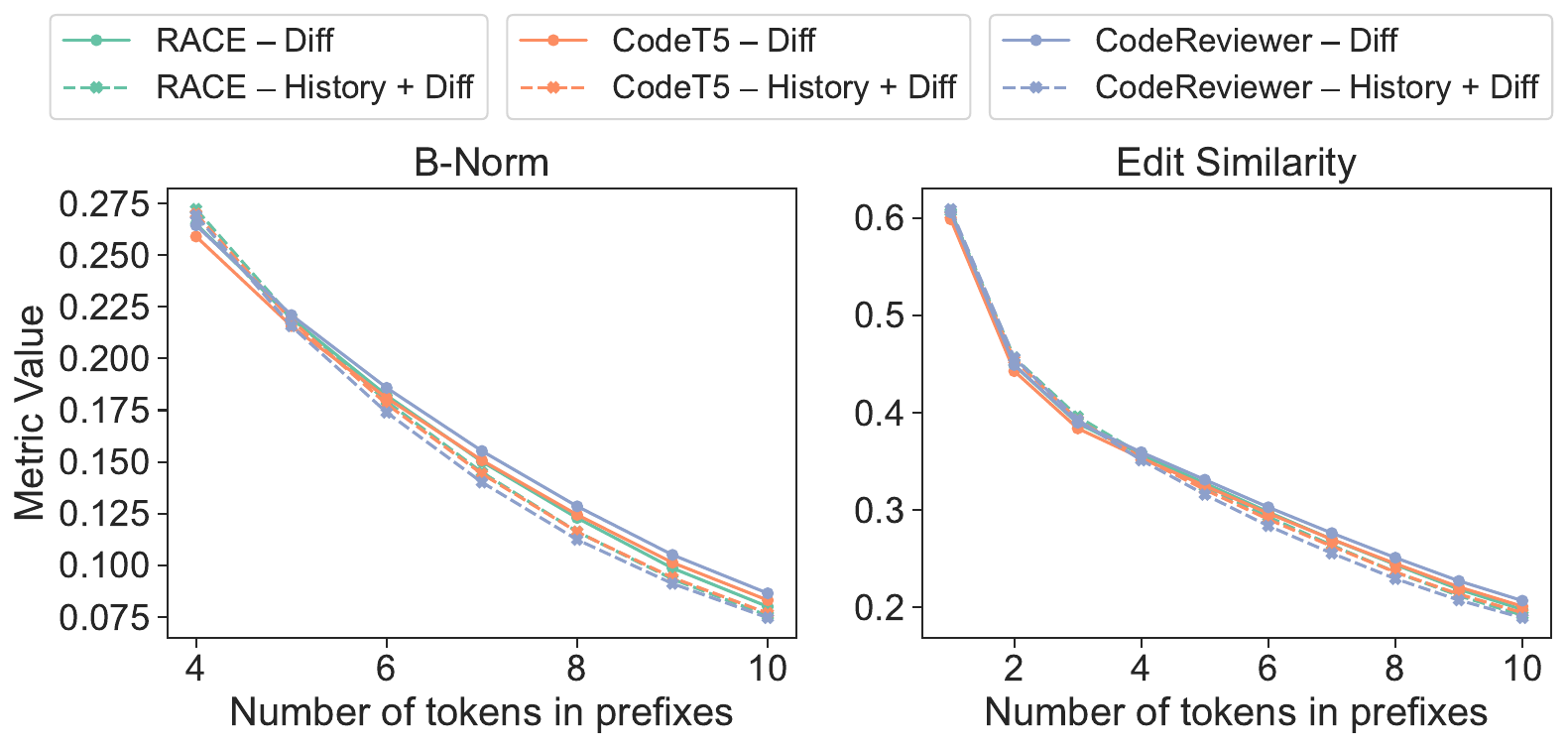}
    \vspace{-0.5cm}
    \caption{Metrics between prefixes on $CMG_{test}$ for 25\% context ratio (completion setting) for CMG approaches. Dashed lines correspond to models with history.}
    \label{fig:prefix_metrics_25}
     \vspace{-0.3cm}
\end{figure}

Next, we observe that there is but a small difference in absolute metrics values between state-of-the-art CMG approaches, and their ranking is not consistent across different metrics and different context ratios. Additionally, we observe that plots for metrics between prefixes for all the models exhibit very similar patterns --- the models are almost indistinguishable. We share the plots for the 0\% context ratio setting in Figure~\ref{fig:prefix_metrics_0} and the plots for the 25\% context ratio setting in Figure~\ref{fig:prefix_metrics_25}. 

We confirm that the superiority of RACE over both CodeT5 and CodeReviewer is statistically significant in all context ratios. In the 0\% context ratio setting, the difference between CodeT5 and CodeReviewer is not statistically significant, however, CodeReviewer is confirmed to be superior in both completion settings.
This outcome is not completely surprising, since all the models are based on CodeT5, however, previous works suggest significantly larger gaps in performance. For RACE, our setup is different from the original in two aspects: we include more diverse commits in the dataset and we switch from the exact retrieval implementation to the approximate one. As for CodeReviewer, perhaps, the fact that it was intended for the code review tasks plays a larger role, since code review data can be different from commit messages. 

Finally, from Figure~\ref{fig:prefix_metrics_25} we observe that all the metrics values in the completion setup decrease with the prediction length, which is consistent with the work on comment completion~\cite{mastropaolo2021empirical}. However, for generation, from Figure~\ref{fig:prefix_metrics_0} we note that \textit{Edit Similarity} for short sequences is worse than for the sequences of average length. We investigate the examples where \textit{Edit Similarity} for sequences of 4 tokens is higher than \textit{Edit Similarity} for sequences of 2 tokens. We find that many cases are due to conventions, which suggest special prefixes for commit messages, \eg Conventional Commits~\cite{cc}. Consider this example: the ground truth commit message is \texttt{feat: add hideSampleTab option}, and the corresponding prediction is \texttt{Add hideSampleTab option}. CMG approaches can fail to correctly determine the convention for the particular case. This finding highlights the variety of writing styles and conventions for commit messages that exist in the wild. Since in the completion setup all the metrics decrease, this problem may be less relevant to completion. 

\vspace{0.1cm}

\observation{\textbf{Summary of RQ A1.} The completion task is easier than generation. All CMG approaches exhibit similar patterns, but RACE is better than both CodeT5 and CodeReviewer. For completion, performance consistently decreases as the number of tokens to predict grows. For generation, there is an additional challenge of correctly generating the beginning of the sequence.}

\vspace{0.1cm}

\begin{figure}
    \centering
    \includegraphics[width=0.49\textwidth]{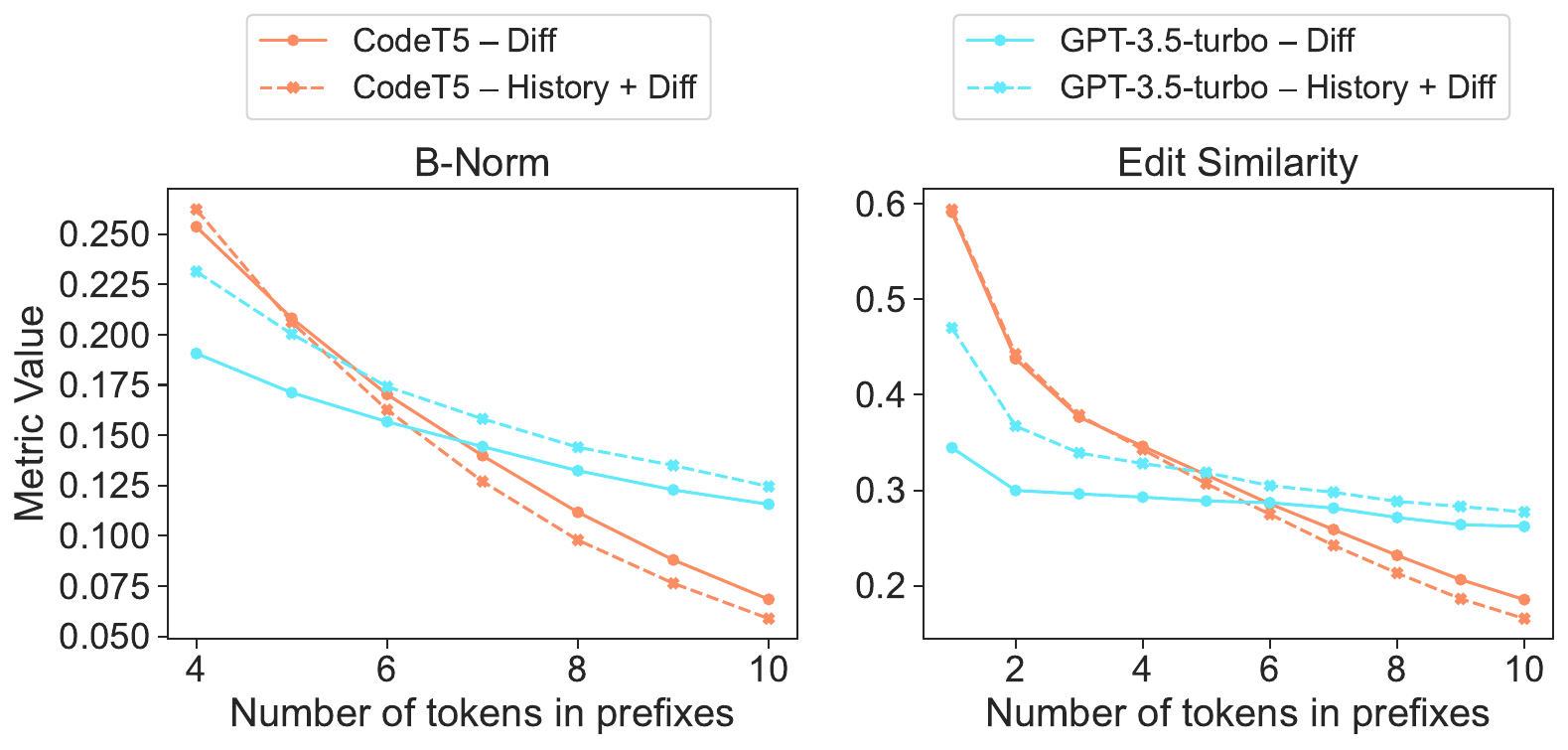}
    \caption{Metrics between prefixes on $LLM_{test}$ for 25\% context ratio (completion setting) for CodeT5 and GPT-3.5-turbo. Dashed lines correspond to models with history.}
    \label{fig:llm_prefix_metrics_25}
    \vspace{-0.4cm}
\end{figure}

\textbf{RQ A2: How do LLMs perform in comparison with state-of-the-art CMG approaches?}
To answer this research question, we obtain predictions from GPT-3.5-turbo on the $LLM_{test}$ dataset. Since $LLM_{test}$ is a subset of $CMG_{test}$, we also recalculate the metrics for all CMG approaches using only predictions for $LLM_{test}$. We present the results in Table~\ref{tab:llm-full-metrics}. In Figure~\ref{fig:llm_prefix_metrics_25}, we also present  plots for metrics between prefixes on $LLM_{test}$. Similar to $CMG_{test}$, different CMG approaches exhibit very similar patterns, so we only include metrics for CodeT5 for clarity. Due to the lack of space, we only include the results for the 25\% context ratio. We note that our findings mostly hold true for the remaining settings as well. We provide the rest of the results in our online appendix~\cite{artifacts}.

\input{tables/07-2-CMG-LLM}

While the results of the CMG approaches on $LLM_{test}$ are slightly different from the results on a larger dataset $CMG_{test}$, they remain of the same magnitude, and most patterns still hold true. Hence, we consider $LLM_{test}$ a reliable dataset to estimate LLM performance.
We observe that the metrics values between full predictions and targets are lower for GPT-3.5-turbo (line 8) than for the CMG approaches (lines 4--6). We confirm that the difference is statistically significant.

Figure~\ref{fig:llm_prefix_metrics_25} shows that while GPT-3.5-turbo is worse than CodeT5 for short sequences, it becomes better as the number of tokens to predict grows. By empirically investigating predictions, we note that GPT-3.5-turbo often tries to produce detailed messages and completely exhausts the maximum tokens restriction, while CMG models often stop early. In the $25\%$ context ratio setting, the median number of tokens for GPT-3.5-turbo and for all CMG models is $9$ and $4$, respectively. Additionally, we observe that GPT-3.5-turbo may disregard the given prefixes and generate novel messages instead. With CMG approaches, we explicitly restrict the beam search procedure to consider only hypotheses starting from the given prefixes, which is not possible for the LLM available through the API. This might be the underlying reason for the CMG approaches being especially superior in terms of metrics between short prefixes and \textit{ExactMatch}.

\observation{\textbf{Summary of RQ A2.} Generally, and especially for short sequences, GPT-3.5-turbo is worse than CMG approaches. However, it may be a better choice for generating long and detailed messages. Moreover, we note that our LLM setting is very simple, and further investigation is required to uncover the full potential of LLMs in commit message generation and completion tasks.
}


\subsection{History and Diversity of Data}

\textbf{RQ B1: How does using commit message history as an additional input affect the models’ quality?}
From Table~\ref{tab:main-full-metrics}, we observe that \textit{B-Norm} and \textit{Edit Similarity} metrics across all the models and settings increase when adding history. In the generation setting, \textit{ExactMatch} is drastically better for the models with history (lines 1--3) compared to the models without history (lines 4--6). However, for completion, history slightly distracts the models (lines 7--9 with history and lines 10--12 without history, 25\% context ratio).
We confirm that the impact of history in terms of \textit{B-Norm} and \textit{Edit Similarity} is statistically significant, as well as for \textit{ExactMatch} in the generation setting. However, not all the models exhibit statistically significant difference in \textit{ExactMatch} in completion settings.

From Figure~\ref{fig:prefix_metrics_0} with the plots with metrics between prefixes in the generation setting, we observe that history improves the results on short sequences. Hence, commit message history might be helpful to determine suitable conventions for the current author and repository when generating messages from scratch. However, plots with metrics between prefixes in the completion setting from Figure~\ref{fig:prefix_metrics_25} show that adding the history barely impacts the completion of a given message prefix: the plots for the models with and without history follow similar patterns, and the models with history are even worse in some cases. This might indicate that the beginning of the commit message is extra challenging to generate correctly. Since the beginning of the commit message is given in the completion setting, the commit message history brings little value.

Apart from the CMG approaches, we also experiment with providing GPT-3.5-turbo a previous message from the commit message history. We observe that GPT-3.5-turbo benefits from history (lines 7-8 from Table~\ref{tab:llm-full-metrics}) and we confirm that the difference is statistically significant.

\observation{\textbf{Summary  of RQ B1.} Our approach of integrating commit message history into CMG approaches turned out to be useful for generation, but the results for completion are conflicting. In contrast, GPT-3.5-turbo benefits from the history both in the completion and generation setting. Overall, commit message history holds potential, but more exploration is necessary.}

\textbf{RQ B2: How do state-of-the-art CMG approaches perform with and without common data filtering steps?}
To answer this research question, we study the most frequent filters, namely, \textit{First Sentence}, \textit{Verb-Direct Object}, and \textit{Diff Length $\leq 100$ tokens}. We do not consider \textit{Message Length $\leq 30$ tokens}, because all the models are allowed to generate only up to $15$ tokens, so naturally the metrics for the subset where all the targets are longer would be low.

In our $CMG_{test}$ dataset, $10,385$ examples ($5.08\%$) fit all the filters and $22,332$ ($10.93\%$) examples fit neither of the filters. In Table~\ref{tab:filter-full-metrics}, we present the metrics for the Filtered subset, the Out-of-Filters subset of $10,385$ examples (fitting \emph{no} filters), and a random subset of $10,385$ examples. Due to the lack of space, we only include the results for the 25\% context ratio. We note that our findings mostly hold true for the remaining settings as well. We provide the rest of the results in our online appendix~\cite{artifacts}.

\input{tables/07-3-filters}

We observe from Table~\ref{tab:filter-full-metrics} that all CMG approaches achieve higher results on the Filtered subset (lines 4--6) than on the Random subset (lines 10--12), and the results on the Out-of-Filters subset (lines 16--18) are by far the lowest (except for \textit{ExactMatch}). The difference is statistically significant for all the models, with an exception for \textit{ExactMatch} for some cases. Therefore, using the Filtered subset may not provide a reliable estimate of the quality on a diverse set of commits. Moreover, generation or completion for Out-of-Filters commits is more challenging, and current approaches seem to struggle with this task. Note that the data in the Out-of-Filters subset was subject to cleaning as described in Section~\ref{sec:Dataset}.  

We also check whether any of our previous findings about commit message history differ when we consider only Filtered or only Out-of-Filters subsets. In Table~\ref{tab:filter-full-metrics}, when using commit message history together with diffs on the Filtered subset (lines 1--3), CMG approaches perform better than when only using diffs (lines 4--6). These results are statistically significant. Moreover, we observe that the difference between the models with and without history tends to be higher for the Filtered subset than it is for the Random. For example, consider RACE: the increase in \textit{B-Norm} on the Filtered subset is $25.17\%$ (from $23.48$ to $29.39$), while on the Random subset, it is $21.23\%$ (from $18.28$ to  $22.16$). In contrast, on the Out-of-filters subset, additional history input decreases the performance of the CMG approaches. In completion settings, these results are statistically significant. These results may indicate that choosing the correct conventions is not the only challenge for commits that do not fit the common filters or that our method of integrating commit message history into CMG approaches is not tailored for this specific case.

\observation{\textbf{Summary of RQ B2.} For the Out-of-Filters subset, the inclusion of history does not improve the results, which requires additional exploration. The average results of CMG approaches are different from both Filtered and Out-of-Filters subsets --- Filtered results are better, while Out-of-Filters results are worse. Hence, the usage of restrictive filtering makes the evaluation results less representative.
}

\section{Related work}
\label{sec:rw}

A question of the practical applicability of existing commit message generation approaches was raised in the recent study~\cite{qacom}. In several studies, over 50\% of generated messages were rated as low-quality by human experts~\cite{Jiang2017, nngen, qacom}. As a way to overcome this issue, Wang et al.~\cite{qacom} propose a  quality assurance framework \textit{QACom} that determines whether a given message is semantically relevant for a given diff. It can be used on top of any CMG approach to avoid showing low-quality predictions to users. 

In our work, we propose to explore another option --- repurposing commit message generation approaches for the completion task. Researchers have already tackled the completion of code comments~\cite{Ciurumelea2020SuggestingCC, mastropaolo2021empirical}. Mastropaolo et al.~\cite{mastropaolo2021empirical} explored the performance of a simple n-gram model that uses only the prefix of a comment typed by the developer as an input, as well as a Text-To-Text Transfer Transformer (T5)~\cite{t5} that utilizes both the prefix and the related code context. Commit messages are similar to code comments in the sense that both are natural language artifacts for software maintenance and comprehension. However, both domains have their unique challenges. 
To the best of our knowledge, there is no published work on the completion of commit messages.

\section{Threats to Validity}
\label{sec:ttc}

\textbf{Internal validity}. We identified and addressed these threats:

\textit{Data quality}. Since we collect a large number of commits from open-source repositories, their quality is a potential threat to validity. To mitigate this threat, we followed best practices from previous work to clean the data (\eg deduplicate and remove messages from bots) and extensively described our processing pipeline. 

 \textit{Hyperparameters}. Since we consider several approaches and settings and train them on a large-scale dataset, it is infeasible to tune optimal hyperparameters for each individual approach. To mitigate this threat, we employed hyperparameters setting from a previous study by Shi et al.~\cite{race}.
 
 \textit{Implementation Errors}. The presence of bugs in the source code implemented for this study poses a potential threat to validity. To mitigate this threat, we thoroughly tested the code and followed software engineering best practices to ensure that it was written in a clear and organized manner. Despite our efforts, there is still a possibility of undetected errors being present in the implementation.

\textbf{External validity.} We identified the following threats:

 \textit{Model selection}. We included several CMG approaches and a recent general-purpose LLM in our experiments. Still, our findings may not generalize to the models that were not considered in this study.
 
 \textit{Dataset}. Our findings may not transfer to different kinds of commits that were not present in our dataset, for instance, commits in other programming languages (beyond the 20 included in our dataset) or commits from proprietary codebases.

Even though these threats are important to note, we believe that they do not invalidate the main contributions of this work and the important novel ideas that we evaluated.

\textbf{Verifiability.}
We provide all the necessary details about our study to help others
replicate. We released publicly the data~\cite{dataset} and all the code for our models and experiments~\cite{artifacts}.

\section{Future Work}
\label{sec:future-work}

Our future work aims to expand upon the two ideas presented in this study. In particular, we only considered a single way of integrating commit message history, and other approaches may result in further improvements. Moreover, our approach implies extending models' inputs with previous commit messages. All of the models in our experiments follow the Transformer architecture~\cite{transformer}, whose time complexity scales quadratically with the number of tokens in inputs, so longer contexts might harm the performance. Hence, a promising direction might be to develop a separate model for producing efficient representations of developers' writing style or project conventions based on commit message history. Likewise, we only investigated a single zero-shot setting for a single LLM, and other LLMs or more advanced prompt engineering techniques are worth exploring.

Another important research direction is semantic evaluation. We followed best practices by using the B-Norm metric~\cite{mcmd}, however, all the metrics we employed are based on the overlap of words or characters between the generated and the reference messages, which may not fully reflect the semantic quality aspects, such as adequacy or usefulness. Human evaluation might uncover new insights for all the RQs addressed in our work. Alternatively, it could be intriguing to investigate the applicability of the recent automatic semantic metrics, including using LLMs as evaluators~\cite{llm-as-a-judge}, for CMG task.

Finally, expanding the scope to diverse commits brings novel challenges. Firstly, there is a possible concern of data quality, especially regarding the commit messages. Considering that the notion of commit message quality gained attention in several recent studies~\cite{what_makes_a_good_msg, icse2023quality},  additional filtering of our dataset might be required. 
Secondly, we believe that conducting a deeper analysis of the commits that current CMG approaches struggle with could benefit the field.

\section{Conclusions}
\label{sec:conclusion}

This work sheds light on the potential of two novel ideas for personalized commit message generation, namely, \textit{focusing on the completion task} and \textit{integrating commit message history as context}. We conduct experiments with several CMG approaches, including previously proposed models from the literature, and a recent LLM GPT-3.5-turbo. Our results suggest that both ideas show promise when implemented independently, however, yield questionable improvements when implemented together. Furthermore, we highlight that several data filtering steps employed in previous works are overly restrictive and exclude many real-world examples. Based on our findings, relying solely on the results obtained on filtered commits might not provide a reliable estimate of the overall performance. In contrast, commits that fall outside of the common filters present new challenges to existing CMG approaches. Finally, we show that the overall quality of GPT-3.5-turbo in the zero-shot setting is inferior to existing state-of-the-art CMG approaches, but it has the potential for generating detailed commit messages. 

\bibliographystyle{IEEEtran}
\balance
\bibliography{IEEEabrv, sample-base}

\end{document}

%% file: tables/04-datasets-statistics.tex
\begin{table*}[h]
        \caption{Statistics on restrictive filters in existing CMG datasets.}
    \begin{tabular}{ccccccc}
        \toprule
         \multirow{2}{*}{\textbf{CMG Dataset}} & \multicolumn{3}{c}{\textbf{Commit message filters}} & \multicolumn{3}{c}{\textbf{Commit diff filters}}\\ \cmidrule(lr){2-4}\cmidrule(lr){5-7} 
          & \textbf{First Sentence} & \textbf{Message Structure} & \textbf{Message Length} & \textbf{Only Code} & \textbf{Diff Length (\# tokens)} & \textbf{Diff Length (other)} \\
        \midrule
        Loyola et al.~\cite{loyola2017neural} & -- & -- & -- & -- & --& --\\
        Jiang et al.~\cite{Jiang2017}& + & V-DO & $\leq 30$ tokens & -- & $\leq 100$ tokens&--\\
        NNGen.~\cite{nngen} & + & V-DO & $\leq 30$ tokens& -- & $\leq 100$ tokens&--\\
        PtrGNCMsg~\cite{ptrmsg} & + & V-DO & $\leq 30$ tokens& -- & $\leq 100$ tokens&--\\
        CoDiSum~\cite{codisum} & -- & -- & $\leq 20$ tokens& + & $\leq 200$ tokens &-- \\
        CoReC~\cite{corec} & + & V-DO & $\leq 30$ tokens& -- & $\leq 100$ tokens&--\\
        ATOM~\cite{atom} & -- & -- & $\leq 20$ tokens& + & -- &$\leq 5$ chunks\\
        MCMD~\cite{mcmd} & + & -- & --& --& --&--\\
        CommitBERT~\cite{commitbert} & + & Verbs &--& + &-- &$\leq 2$ files\\
        \midrule
        \textbf{Total} & 6/9 & 5/9 & 6/9 & 3/9 & 5/9 & 2/9\\
        \bottomrule
    \end{tabular}
    \vspace{0.3cm}
\label{tab:filters3}
    \vspace{-0.5cm}
\end{table*}

%% file: tables/05-our-dataset.tex
\begin{table}[t]
    \centering
    \caption{Number of commits in the obtained CommitChronicle dataset for each programming language. Languages are sorted by the total number of corresponding commits.}
    \begin{tabular}{c c c c c}
    \toprule
    \textbf{Language} & \textbf{Train} & \textbf{Validation} & \textbf{Test} & \textbf{Total}\\
    \midrule
    Python & 1,330,155 &  212,563 &  247,421 &  1,790,139 \\
 JavaScript & 1,076,877 &  169,502 &  229,720 &  1,476,099 \\
       Java &  952,162 &  200,035 &  204,862 &  1,357,059 \\
 TypeScript &  936,697 &  177,258 &  198,272 &  1,312,227 \\
        C++ &  830,683 &  201,716 &  123,725 &  1,156,124 \\
         Go &  672,045 &  133,954 &  134,699 &   940,698 \\
         C\# &  425,642 &   84,708 &   65,528 &   575,878 \\
          C &  309,153 &   57,970 &   38,340 &   405,463 \\
       Rust &  240,037 &   60,788 &   45,167 &   345,992 \\
       Ruby &  181,916 &   39,912 &   33,433 &   255,261 \\
        PHP &  178,556 &   32,293 &   36,618 &   247,467 \\
     Kotlin &  154,276 &   29,781 &   28,021 &   212,078 \\
      Shell &  117,927 &   13,902 &   27,019 &   158,848 \\
      Swift &  101,274 &   28,227 &    8,938 &   138,439 \\
        Nix &    2,526 &   86,022 &    8,108 &    96,656 \\
     Groovy &   23,262 &    1,745 &   38,799 &    63,806 \\
       Dart &   42,061 &   17,527 &    2,895 &    62,483 \\
     Elixir &   41,562 &    3,380 &    5,874 &    50,816 \\
Objective-C &   32,517 &    1,294 &    7,708 &    41,519 \\
  Smalltalk &   10,130 &    1,465 &    1,120 &    12,715 \\
    \midrule
      Total & 7,659,458 & 1,554,042 & 1,486,267 & 10,699,767 \\
    \bottomrule
    \end{tabular}
    \vspace{0.3cm}

    \vspace{-0.7cm}
        \label{tab:final_statistics}
\end{table}

%% file: tables/07-1-generation-completion.tex
\begin{table}
    \centering
    \caption{Results for CMG approaches on $CMG_{test}$. \textit{EdSim} stands for \textit{Edit Similarity}, \textit{EM} stands for \textit{ExactMatch}. \textit{CoRev} is the CodeReviewer model~\cite{codereviewer}. Percentages represent the length of the provided message prefix, with 0\% being the generation task.}\label{tab:main-full-metrics}
    \begin{tabular}{cl
                    cccc|c}
        \toprule
        &\multicolumn{1}{c}{\textbf{Approach}} & 
        \textbf{B-Norm} & 
        \textbf{EdSim} & 
        \textbf{EM@1} &
        \textbf{EM@2} &
        \textcolor{gray}{\textbf{\textnumero}}\\
        \midrule
        \multirow{6}{*}{\rotatebox[origin=c]{90}{\textbf{0\% (Gen.)}}} & CodeT5, history & 
        $16.80$ & $30.91$ & $17.68$ & $4.27$ & \textcolor{gray}{1}\\
        &RACE, history & 
        \textbf{16.91} & \textbf{31.15} & \textbf{17.95} & $4.36$ & \textcolor{gray}{2}\\
        &CoRev, history & 
        $16.78$ & $30.74$ & $17.83$ & \textbf{4.38} & \textcolor{gray}{3}\\
        \cmidrule{2-7}
        &CodeT5 & 
        $15.12$ & $28.71$ & $10.90$ & $3.03$& \textcolor{gray}{4}\\
       & RACE & 
        \textbf{15.32} & \textbf{29.02} & \textbf{11.37} & \textbf{3.07} & \textcolor{gray}{5}\\
        &CoRev & 
        $15.15$ & $28.76$ & $10.87$ & $3.05$& \textcolor{gray}{6}\\
        \midrule
        \multirow{6}{*}{\rotatebox[origin=c]{90}{\textbf{25\% (Compl.)}}} & CodeT5, history & 
        $21.94$  & $33.31$ & $44.98$ & $13.10$& \textcolor{gray}{7}\\
        &RACE, history & 
        \textbf{22.16}  & \textbf{33.78} & $45.36$ & \textbf{13.40} & \textcolor{gray}{8}\\
        &CoRev, history & 
        $21.84$ & $32.90$ & \textbf{45.58} & $13.28$& \textcolor{gray}{9}\\
        \cmidrule{2-7}
        &CodeT5 & 
        $17.91$  & $30.54$ & $45.35$ & $12.92$& \textcolor{gray}{10}\\
        &RACE & 
        \textbf{18.38}  & $30.91$ & \textbf{46.62} & \textbf{13.45}& \textcolor{gray}{11}\\
       & CoRev & 
        $18.10$ & \textbf{30.93} & $46.05$ & $13.35$& \textcolor{gray}{12}\\
        \midrule
        \multirow{6}{*}{\rotatebox[origin=c]{90}{\textbf{50\% (Compl.)}}} & CodeT5, history & 
        $26.90$ & $33.95$ & $47.45$ & $12.75$& \textcolor{gray}{13}\\
        &RACE, history & 
        \textbf{27.28}  & \textbf{34.69} & $47.84$ & \textbf{13.26} & \textcolor{gray}{14}\\
       & CoRev, history & 
        $26.94$  & $33.76$ & \textbf{48.10} & $12.90$& \textcolor{gray}{15}\\
        \cmidrule{2-7}
        &CodeT5 & 
        $24.13$  & $32.74$ & $49.94$ & $14.03$& \textcolor{gray}{16}\\
       & RACE & 
        \textbf{24.74}  & \textbf{33.22} & $50.68$ & $14.38$& \textcolor{gray}{17}\\
        &CoRev & 
        $24.35$  & $33.20$ & \textbf{50.90} & \textbf{14.59}& \textcolor{gray}{18}\\
        \bottomrule
    \end{tabular}
    \vspace{0.3cm}
    
    \vspace{-0.7cm}
\end{table}

%% file: tables/07-2-CMG-LLM.tex
\begin{table}
\caption{Results for CMG approaches and GPT-3.5-turbo on $LLM_{test}$. All the results are presented for the $25\%$ context ratio (completion setting). \textit{EdSim} stands for \textit{Edit Similarity}, \textit{EM} stands for \textit{ExactMatch}.  \textit{CoRev} is the CodeReviewer model~\cite{codereviewer}.}\label{tab:llm-full-metrics}
    \centering
    \begin{tabular}{l
                    cccc|c}
        \toprule
        \multicolumn{1}{c}{\textbf{Approach}} & 
        \textbf{B-Norm} & 
        \textbf{EdSim} & 
        \textbf{EM@1} &
        \textbf{EM@2}&
        \textcolor{gray}{\textbf{\textnumero}}\\
        \midrule
        CodeT5, history & 
        $21.11$ & $32.31$ & $43.68$ & $12.45$&\textcolor{gray}{1}\\
        RACE, history & 
        $21.14$ & $32.35$ & $44.92$ & $12.93$&\textcolor{gray}{2}\\
        CoRev, history & 
        \textbf{21.35} & \textbf{32.68} & \textbf{45.69} & \textbf{13.15} &\textcolor{gray}{3}\\
        \cmidrule{1-6}
        CodeT5 & 
        $17.16$ & $30.02$ & $45.19$ & $12.85$&\textcolor{gray}{4}\\
        RACE & 
        \textbf{17.54} & $30.13$ & \textbf{46.68} & \textbf{13.33} &\textcolor{gray}{5}\\
        CoRev & 
        $17.34$ & \textbf{30.45} & $45.96$ & $13.03$&\textcolor{gray}{6}\\
        \midrule
        GPT-3.5-turbo, history &
        \textbf{13.24} & \textbf{27.83} & \textbf{34.34} & \textbf{10.09} &\textcolor{gray}{7}\\
        GPT-3.5-turbo & 
        $11.48$ & $26.35$ & $21.84$ & $5.99$&\textcolor{gray}{8}\\
        \bottomrule
    \end{tabular}
    \vspace{-0.2cm}
    
\end{table}

%% file: tables/07-3-filters.tex
\begin{table}
\caption{Results for CMG approaches on filtered, random, and out-of-filters subsets of $CMG_{test}$ with $10,385$ examples. All the results are presented for the 25\% context ratio (completion setting). \textit{EdSim} stands for \textit{Edit Similarity}, \textit{EM} stands for \textit{ExactMatch}. \textit{CoRev} is the CodeReviewer model~\cite{codereviewer}.}\label{tab:filter-full-metrics}
    \centering
    \begin{tabular}{cl
                    cccc|c}
        \toprule
        & 
        \multicolumn{1}{c}{\textbf{Approach}} & 
        \textbf{B-Norm} & 
        \textbf{EdSim} & 
        \textbf{EM@1} &
        \textbf{EM@2}&
        \textcolor{gray}{\textbf{\textnumero}}\\
        \midrule
        \multirow{6}{*}{\rotatebox[origin=c]{90}{\textbf{Filtered}}} 
        & CodeT5, history & 
        $29.21$ & $39.36$ & $51.55$ & $16.29$&\textcolor{gray}{1}\\
        & RACE, history & 
        \textbf{29.39}  & \textbf{40.25} & \textbf{52.35} & \textbf{17.00} &\textcolor{gray}{2}\\
        & CoRev, history & 
        $29.22$ & $39.35$ & $51.99$ & $16.89$&\textcolor{gray}{3}\\
        \cmidrule{2-7}
        & CodeT5 & 
        $22.77$ & $35.11$ & $48.51$ & $15.55$&\textcolor{gray}{4}\\
        & RACE & 
        \textbf{23.48} & \textbf{35.85} & \textbf{49.91} & \textbf{16.28} &\textcolor{gray}{5}\\
        & CoRev & 
        $22.88$ & $35.56$ & $49.04$ & $15.67$&\textcolor{gray}{6}\\
        \midrule
        \multirow{6}{*}{\rotatebox[origin=c]{90}{\textbf{Random}}}
        & CodeT5, history & 
        $21.74$ & $33.24$ & $44.26$ & $13.43$&\textcolor{gray}{7}\\
        & RACE, history & 
        \textbf{22.16} & \textbf{33.81} & \textbf{45.12} & \textbf{13.72} &\textcolor{gray}{8}\\
        & CoRev, history & 
        $21.71$ & $32.92$ & $45.11$ & $13.42$&\textcolor{gray}{9}\\
        \cmidrule{2-7}
        & CodeT5 & 
        $17.66$ & $30.48$ & $44.88$ & $12.93$&\textcolor{gray}{10}\\
        & RACE & 
        \textbf{18.28} & \textbf{30.95} & \textbf{46.24} & \textbf{14.05} &\textcolor{gray}{11}\\
        & CoRev & 
        $18.04$ & $30.94$ & $45.51$ & $13.82$&\textcolor{gray}{12}\\
        \midrule
       \multirow{6}{*}{\rotatebox[origin=c]{90}{\textbf{Out-of-filters}}}
        & CodeT5, history & 
        \textbf{5.61} & \textbf{16.00} & $44.55$ & \textbf{10.52}&\textcolor{gray}{13}\\
        & RACE, history & 
        $5.52$ & $15.54$ & $44.48$ & $10.02$&\textcolor{gray}{14}\\
        & CoRev, history & 
        $5.41$ & $15.42$ & \textbf{45.38} & $10.40$&\textcolor{gray}{15}\\
        \cmidrule{2-7}
        & CodeT5 & 
        $6.21$ & $16.44$ & $47.41$ & $10.98$&\textcolor{gray}{16}\\
        & RACE & 
        $6.16$ & $16.21$ & $47.68$ & $11.47$&\textcolor{gray}{17}\\
        & CoRev & 
        \textbf{6.37} & \textbf{16.71} & \textbf{48.79} & \textbf{11.64} &\textcolor{gray}{18}\\
        \bottomrule
    \end{tabular}
    \vspace{-0.3cm}
    
\end{table}